\documentclass{PoS}

\usepackage{colordvi}
\usepackage{epsfig}
\usepackage{axodraw}
\usepackage{epsfig}
\usepackage{graphicx}
\usepackage{rotate}
\usepackage{latexsym}
\usepackage{amssymb}
\usepackage{amsmath}
\usepackage{multirow}
\usepackage{dsfont}

\def\refeq#1{\mbox{Eq.~(\ref{#1})}}

\def\citere#1{\mbox{Ref.~\cite{#1}}}

\def\beq{\begin{equation}}
\def\eeq{\end{equation}}
\def\beqar{\begin{eqnarray}}
\def\eeqar{\end{eqnarray}}

\def\ie{i.e.\ }

\newcommand{\ps}{p\hspace{-0.42em}/}

\newcommand{\gs}{g_{\mathrm{s}}}
\newcommand{\NC}{N_{\mathrm{c}}}

\def\mathswitch#1{\relax\ifmmode#1\else$#1$\fi}
\def\mathswitchr#1{\relax\ifmmode{\mathrm{#1}}\else$\mathrm{#1}$\fi}
\def\mathswitchit#1{\relax\ifmmode{#1}\else$#1$\fi}

\newcommand{\ri}{{\mathrm{i}}}
\newcommand{\rd}{{\mathrm{d}}}

\newcommand{\recola}{{\sc Recola}}

\title{EW and QCD One-Loop Amplitudes with RECOLA}

\ShortTitle{EW and QCD One-Loop Amplitudes with RECOLA}

\author{S.~Actis \\
        Paul Scherrer Institut\thanks{Former affiliation} \\
        }

\author{A.~Denner \\
        Universit\"at W\"urzburg\\ 
        E-mail: \email{denner@physik.uni-wuerzburg.de}
        }

\author{L.~Hofer\thanks{Present address: Universitat Aut\`onoma de Barcelona} \\
        Universit\"at W\"urzburg\\
        E-mail: \email{lhofer@ifae.es}
        }

\author{A.~Scharf \\
        Universit\"at W\"urzburg\\
        E-mail: \email{ascharf@physik.uni-wuerzburg.de}
        }

\author{\speaker{S.~Uccirati}
        \\
        Torino University and INFN \\
        E-mail: \email{uccirati@to.infn.it}
        }

\abstract{
We present the computer code \recola\ for the computation of EW and 
QCD amplitudes in the Standard Model at next-to-leading order. 
One-loop amplitudes are represented as linear combinations of tensor 
integrals whose coefficients are calculated by means of recursive 
relations similar to Dyson-Schwinger equations.
A novel treatment of colour enables us to recursively 
construct the colour structure of the amplitude efficiently.  
\recola\ is linked with the library COLLIER for the computation of 
the tensor integrals.
}

\FullConference{
11th International Symposium on Radiative Corrections 
(Applications of Quantum Field Theory to Phenomenology) 
(RADCOR 2013),\\
22-27 September 2013\\
Lumley Castle Hotel, Durham, UK}

\begin{document}

\section{Introduction}

\vspace{-.2cm}
All the particles predicted by the Standard Model have been confirmed 
by experiments, yet many sectors of this successful theory have to 
be precisely investigated. 
The description of high energy processes is based on pertubation 
theory, and accurate theoretical predictions require a detailed 
understanding beyond Born approximation. 
At a hadron collider as the LHC, QCD corrections are known to be 
large, but also electroweak (EW) corrections can have an important 
impact. 
Indeed the high energies attained by the LHC allow to collect data
in phase-space regions where the effects of logarithms of EW origin
become sizable. 
In particular cases like e.g. Higgs production in vector-boson fusion 
the EW corrections can be of the same order of magnitude as QCD 
corrections~\cite{Ciccolini:2007ec}.
Finally, photon emission creates mass-singular logarithms which 
again lead to large contributions.
Therefore a proper theoretical description of LHC physics requires
next-to-leading-order (NLO) computations of multiparticle processes
(with five, six, or more external legs) in the full SM (including EW
corrections).

While QCD corrections have been calculated in the last 
years for many processes, in the EW sector the situation is less 
satisfactory.
Usually EW one-loop computations are more involved than QCD ones and 
often rely on codes created ``ad hoc'' to compute specific processes, 
whose generalization to other processes is tedious. 
Moreover the computation of NLO corrections for the elementary 
scattering is not the end of the story and many other issues have 
to be considered in hadronic processes (such as multi-channel 
Monte Carlos, a proper treatment of real emission, convolution with 
the parton distribution functions, parton shower, etc.). 
For this reason the creation of general and automatized codes for 
the computation of NLO partonic matrix elements is highly 
desirable.
In the past years many groups have concentrated their efforts to make
such calculations feasible, and a lot of codes have appeared~\cite{
Arnold:2008rz,Berger:2008sj,Badger:2010nx,Hirschi:2011pa,Bevilacqua:2011xh,
Cullen:2011ac,Cascioli:2011va} with a high level of 
automatization and impressive performances, however restricting their 
range of applicability mostly to the QCD sector of the SM.

For this reason we have developed \recola, a generator 
of tree- and one-loop amplitudes in the full SM \cite{Actis:2012qn}. 
The code overcomes the difficulties in automatizing efficiently 
computations based on Feynman diagrams, choosing an alternative approach 
which extends to one-loop amplitudes the known recursion relations 
for tree-level amplitudes. 
The idea, originally proposed by Andreas van Hameren 
in~\citere{vanHameren:2009vq} for gluonic amplitudes, is to write 
one-loop amplitudes as linear combinations of tensor integrals 
and compute recursively the coefficients of such a decomposition.
The tensor integrals themselves are then computed by linking the 
code to external libraries. 

\section{Tree-level recursion relations}
\label{tree}

\vspace{-.2cm}
The tree-level recursive algorithm is inspired by the Dyson-Schwinger
equations~\cite{Dyson:1949ha}, using off-shell currents as basic 
building blocks.
Let us consider a process with $L$ external legs, written in the form
%
%
$P_1 + \dots + P_{L-1} + P_L \,\to\, 0$,
where we select one particle to be the last ($P_L$) while the others 
are called primary.
The off-shell current 

\vspace{-.4cm}
\beq
w(P,{\cal C},B_{\{l_1,\dots,l_n\}})
\quad=\quad
\vcenter{\hbox{
\scalebox{.7}{
\begin{picture}(55,30)(-5,-15)
\Line(20,0)(40,0)
\GCirc(40,0){1.5}{0}
\Line(0,15)(20,0)
\Line(0,-15)(20,0)
\GCirc(20,0){10}{.8}
\DashCArc(20,0)(22,155,205){2}
\Text(-9,-3)[cb]{$n$} 
\Text(42,6)[cb]{$P$} 
\end{picture}
}
}}
\label{current}
\eeq

\noindent
is defined as 
the amplitude made of $n$ of the $L\!-\!1$ primary on-shell particles 
with labels $\{l_1,\dots,l_n\}$ and the off-shell particle $P$ 
(carrying colour labeled by ${\cal C}$).
The off-shellness of the particle $P$ (depicted by a dot) implies 
that its wave function is not included and, for $n>1$, replaced by 
its propagator.
The off-shell current $w$ has a Lorentz structure of a scalar, spinor or 
vector according to the type of particle $P$.
For $n=1$ the off-shell currents of the external legs coincide with 
their wave functions.
%
%
The labels $\{l_1,\dots,l_n\}$ are expressed through one tag number 
$B_{\{l_1,\dots,l_n\}}$, obtained using the binary notation 
of~\citere{Kanaki:2000ey}:
each external particle $P_i$ is labeled by the tag number 
$B_{\{i\}} = 2^{i-1}$ (\ie $B_{\{1\}} = 1,\;B_{\{2\}} = 2,\; 
B_{\{3\}} = 4,\; B_{\{4\}} = 8, \;\dots$), while the tag number of 
the internal currents is simply obtained by summing the tag 
numbers of the external currents contributing to it (for example 
$B_{\{1,2,4\}} = B_{\{1\}}+B_{\{2\}}+B_{\{4\}} = 1+2+8 = 11$). 
%

The practical advantage of using off-shell currents is that they can 
be computed recursively using the Dyson-Schwinger equation, which 
for the SM reads:

\vspace{-.8cm}
\beq
\scalebox{.9}{
\begin{picture}(55,40)(-5,-2)
\Line(20,0)(40,0)
\GCirc(40,0){1.5}{0}
\Line(0,15)(20,0)
\Line(0,-15)(20,0)
\GCirc(20,0){10}{.8}
\DashCArc(20,0)(22,155,205){2}
\Text(-9,-3)[cb]{$n$} 
\Text(42,5)[cb]{$P$} 
\end{picture}
}
\quad
=
\quad
\sum_{\{i\},\{j\}}^{i+j=n}\;
\sum_{P_i,P_j}\;
\scalebox{.9}{
\begin{picture}(65,40)(-10,-2)
\Line(0,25)(15,15)
\Line(0,5)(15,15)
\Line(15,15)(35,0)
\GCirc(15,15){7.5}{.8}
\DashCArc(15,15)(16,160,200){2}
\Text(31,8)[cb]{\scriptsize $P_{\!i}$}
\Text(-6,12)[cb]{$i$} 
\Line(0,-5)(15,-15)
\Line(0,-25)(15,-15)
\Line(15,-15)(35,0)
\GCirc(15,-15){7.5}{.8}
\DashCArc(15,-15)(16,160,200){2}
\Text(31,-17)[cb]{\scriptsize $P_{\!j}$}
\Text(-6,-18)[cb]{$j$} 
\SetWidth{1.5}
\Line(35,0)(50,0)
\SetWidth{.5}
\BBoxc(34,0)(5,5)
\GCirc(50,0){1.5}{0}
\Text(52,5)[cb]{$P$} 
\end{picture}
}
\quad
+
\;
\sum_{\{i\},\{j\},\{k\}}^{i+j+k=n}\;
\sum_{P_{\!i},P_{\!j},P_{\!k}}\;
\scalebox{.9}{
\begin{picture}(70,40)(-10,-2)
\Line(0,35)(15,25)
\Line(0,15)(15,25)
\Line(15,25)(45,0)
\GCirc(15,25){7.5}{.8}
\DashCArc(15,25)(16,160,200){2}
\Text(-6,22)[cb]{$i$} 
\Text(31,18)[cb]{\scriptsize $P_{\!i}$}
\Line(0,10)(15,0)
\Line(0,-10)(15,0)
\Line(15,0)(45,0)
\GCirc(15,0){7.5}{.8}
\DashCArc(15,0)(16,160,200){2}
\Text(-6,-3)[cb]{$j$} 
\Text(27,1)[cb]{\scriptsize $P_{\!j}$}
\Line(0,-15)(15,-25)
\Line(0,-35)(15,-25)
\Line(15,-25)(45,0)
\GCirc(15,-25){7.5}{.8}
\DashCArc(15,-25)(16,160,200){2}
\Text(-6,-28)[cb]{$k$} 
\Text(31,-27)[cb]{\scriptsize $P_{\!k}$}
\SetWidth{1.5}
\Line(45,0)(60,0)
\SetWidth{.5}
\BBoxc(43,0)(5,5)
\GCirc(60,0){1.5}{0}.
\Text(60,5)[cb]{$P$}
\end{picture}
}
\;\;\;.
\label{recursive}
\eeq

\vspace{.45cm}
\noindent
Each term of the sums represents a ``branch", which is obtained by 
multiplying the generating currents with the interaction vertex, 
marked by the small box, and the propagator of $P$, marked by the
thick line.
The recursion procedure consists in using the Dyson-Schwinger 
equation starting from $n=2$, for whatever allowed particle $P$, obtaining
the 2-leg currents:

\vspace{-.5cm}
\beq
\scalebox{.6}{
\begin{picture}(55,30)(-5,-5)
\Line(20,0)(40,0)
\GCirc(40,0){1.5}{0}
\Line(0,15)(20,0)
\Line(0,-15)(20,0)
\GCirc(20,0){10}{.8}
\end{picture}
}
=
\scalebox{.6}{
\begin{picture}(65,30)(-10,-5)
\GCirc(45,0){1.5}{0}
\Line(15,15)(30,0)
\Line(15,-15)(30,0)
\SetWidth{1.5}
\Line(30,0)(45,0)
\SetWidth{.5}
\BBoxc(29,0)(4,4)
\end{picture}
}.
\eeq

\noindent
This result is then used to compute the 3-leg currents, again for all 
possible $P$:

\vspace{-.3cm}
\beq
\scalebox{.6}{
\begin{picture}(55,30)(-5,-5)
\Line(20,0)(40,0)
\GCirc(40,0){1.5}{0}
\Line(0,15)(20,0)
\Line(0,0)(20,0)
\Line(0,-15)(20,0)
\GCirc(20,0){10}{.8}
\end{picture}
}
=
\scalebox{.6}{
\begin{picture}(65,30)(-10,-5)
\GCirc(45,0){1.5}{0}
\Line(0,25)(15,15)
\Line(0,5)(15,15)
\Line(15,15)(30,0)
\GCirc(15,15){7.5}{.8}
\Line(15,-15)(30,0)
\SetWidth{1.5}
\Line(30,0)(45,0)
\SetWidth{.9}
\BBoxc(29,0)(4,4)
\end{picture}
}
+
\scalebox{.6}{
\begin{picture}(65,30)(-10,-5)
\GCirc(45,0){1.5}{0}
\Line(15,15)(30,0)
\Line(0,-5)(15,-15)
\Line(0,-25)(15,-15)
\Line(15,-15)(30,0)
\GCirc(15,-15){7.5}{.8}
\SetWidth{1.5}
\Line(30,0)(45,0)
\SetWidth{.9}
\BBoxc(29,0)(4,4)
\end{picture}
}
+
\scalebox{.6}{
\begin{picture}(65,30)(0,-5)
\GCirc(55,0){1.5}{0}
\Line(15,25)(40,0)
\Line(15,0)(40,0)
\Line(15,-25)(40,0)
\SetWidth{1.5}
\Line(40,0)(55,0)
\SetWidth{.9}
\BBoxc(39,0)(4,4)
\end{picture}
}.
\eeq

\vspace{.15cm}
\noindent
Proceeding in this way, one can compute recursively all currents for 
$n \leq L-1$.
For the last step, i.e. for $n = L-1$, not all currents are needed, 
but just the one with the particle $P$ coinciding with the 
antiparticle $\bar{P}_L$ of the last particle.
The amplitude ${\cal A}$ is then obtained by multiplying this unique 
last current with the inverse of the propagator of $\bar{P}_L$ and with its 
wave function:

\vspace{-.15cm}
\beq
{\cal A} \;=
\vcenter{\hbox{
\scalebox{.8}{
\begin{picture}(80,30)(-35,-15)
\Line(20,0)(40,0)
\GCirc(40,0){1.5}{0}
\Line(0,15)(20,0)
\Line(0,-15)(20,0)
\GCirc(20,0){10}{.8}
\DashCArc(20,0)(22,155,205){2}
\Text(-17,-3)[cb]{$L\!-\!1$} 
\Text(42,5)[cb]{$\bar{P}_L$} 
\end{picture}
}
}}
\times
(\mbox{propagator of } \bar{P}_L)^{-1}
\times\vcenter{\hbox{
\begin{picture}(40,10)(-5,-5)
\Line(0,0)(20,0)
\GCirc(0,0){1.5}{0}
\Text(30,-3)[cb]{$\bar{P}_L$} 
\end{picture}\,\,\,.}}
\label{last step}
\eeq

\vspace{.05cm}
The advantage of the recursion procedure with respect to the usual 
approach based on Feynman diagrams is the possibility to avoid 
recomputing identical sub-graphs contributing to different diagrams; 
furthermore, each current being the sum of many sub-graphs, the 
number of generated objects which are passed to the next step of the 
recursion is reduced with respect to a ``diagram by diagram'' 
procedure.

\section{One-loop recursion relations}
\label{loop}

\vspace{-.2cm}
Every one-loop amplitude can be written as a linear combination of 
tensor integrals (TIs):

\vspace{-.35cm}
\beq
{\cal A} = 
\sum_{t}\,c_{\mu_1\cdots\mu_{r_t}}^{(t)}\,T_{(t)}^{\mu_1\cdots\mu_{r_t}}\,.
\label{tensor splitting}
\eeq

\vspace{-.25cm}
\noindent
The tensor coefficients (TCs) $c_{\mu_1\cdots\mu_{r_t}}^{(t)}$ do not 
depend on the loop momentum $q$, which is present only in the TIs:

\vspace{-.4cm}
\beq
T_{(t)}^{\mu_1\cdots\mu_{r_t}} =
\frac{(2\pi\mu)^{4-D}}{\ri\pi^2}
\int
\rd^D\!q\,\frac{q^{\mu_1}\cdots q^{\mu_{r_t}}}{D_{0}^{(t)}\cdots D_{k_t}^{(t)}},
\qquad
D_{k_t}^{(t)} = (q+p_{k_t}^{(t)})^2 -( m_{k_t}^{(t)})^2.
\eeq

\vspace{-.1cm}
\noindent
The key of an automatized computation of one-loop amplitudes 
for complicated processes is to find an efficient procedure to 
compute the TCs $c_{\mu_1\cdots\mu_{r_t}}^{(t)}$. 
Our approach to solve this problem starts from the work of 
van~Hameren~\cite{vanHameren:2009vq}, which proposed 
a way to generalize the Dyson-Schwinger recursion relation 
to one-loop gluon amplitudes in QCD.
We have further developed this approach to deal with the full SM.  
Thanks to the correspondence between a one-loop diagram with $L$ 
external legs and a tree diagram with $L+2$ external legs (obtained 
by cutting one of the loop lines), one can associate with the loop 
process $P_1 + \dots  + P_L \to 0$ a family of tree processes with 
two more legs $P_1 + \dots  + P_L + P_0 + \bar{P}_0 \,\to\, 0$, 
for any particle $P_0$ of the SM.
This correspondence is however not unique and selection rules have 
to be applied to identify the tree diagram associated with a given 
one-loop diagram. 
The problem basically arises from the fact that the loop diagram can 
be cut at any of its loop lines and that we can run along the loop 
clockwise or counterclockwise. 
As a simple example we consider a three-point function at one loop 
and find six corresponding tree level diagrams:

\vspace{-.9cm}
\beq
\scalebox{.7}{
\begin{picture}(55,30)(-7,-2)
\Line(0,0)(10,0)
\Line(40,-15)(28,-6)
\Line(40,15)(28,6)
\CArc(20,0)(10,0,360)
\Text(-4,-3)[cb]{\small $1$}
\Text(45,-18)[cb]{\small $2$}
\Text(45,12)[cb]{\small $4$}
\end{picture}
}
\;\to\;\;
\scalebox{.7}{
\begin{picture}(55,30)(-7,-2)
\Line(0,0)(10,0)
\Line(40,-15)(28,-6)
\Line(40,15)(28,6)
\CArc(20,0)(10,143,37)
\Line(12,15)(12,6)
\Text(12,12.5)[cb]{\scriptsize $\times$}
\Line(28,15)(28,6)
\Text(28,12.5)[cb]{\scriptsize $\times$}
\Text(-4,-3)[cb]{\small $1$}
\Text(45,-18)[cb]{\small $2$}
\Text(45,12)[cb]{\small $4$}
\Text(12,19)[cb]{\small $8$}
\Text(28,19)[cb]{\small $16$}
\end{picture}
}
+
\scalebox{.7}{
\begin{picture}(55,30)(-7,-2)
\Line(0,0)(10,0)
\Line(40,-15)(28,-6)
\Line(40,15)(28,6)
\CArc(20,0)(10,143,37)
\Line(12,15)(12,6)
\Text(12,12.5)[cb]{\scriptsize $\times$}
\Line(28,15)(28,6)
\Text(28,12.5)[cb]{\scriptsize $\times$}
\Text(-4,-3)[cb]{\small $1$}
\Text(45,-18)[cb]{\small $4$}
\Text(45,12)[cb]{\small $2$}
\Text(12,19)[cb]{\small $8$}
\Text(28,19)[cb]{\small $16$}
\end{picture}
}
+
\scalebox{.7}{
\begin{picture}(55,30)(-7,-2)
\Line(0,0)(10,0)
\Line(40,-15)(28,-6)
\Line(40,15)(28,6)
\CArc(20,0)(10,143,37)
\Line(12,15)(12,6)
\Text(12,12.5)[cb]{\scriptsize $\times$}
\Line(28,15)(28,6)
\Text(28,12.5)[cb]{\scriptsize $\times$}
\Text(-4,-3)[cb]{\small $2$}
\Text(45,-18)[cb]{\small $1$}
\Text(45,12)[cb]{\small $4$}
\Text(12,19)[cb]{\small $8$}
\Text(28,19)[cb]{\small $16$}
\end{picture}
}
+
\scalebox{.7}{
\begin{picture}(55,30)(-7,-2)
\Line(0,0)(10,0)
\Line(40,-15)(28,-6)
\Line(40,15)(28,6)
\CArc(20,0)(10,143,37)
\Line(12,15)(12,6)
\Text(12,12.5)[cb]{\scriptsize $\times$}
\Line(28,15)(28,6)
\Text(28,12.5)[cb]{\scriptsize $\times$}
\Text(-4,-3)[cb]{\small $2$}
\Text(45,-18)[cb]{\small $4$}
\Text(45,12)[cb]{\small $1$}
\Text(12,19)[cb]{\small $8$}
\Text(28,19)[cb]{\small $16$}
\end{picture}
}
+
\scalebox{.7}{
\begin{picture}(55,30)(-7,-2)
\Line(0,0)(10,0)
\Line(40,-15)(28,-6)
\Line(40,15)(28,6)
\CArc(20,0)(10,143,37)
\Line(12,15)(12,6)
\Text(12,12.5)[cb]{\scriptsize $\times$}
\Line(28,15)(28,6)
\Text(28,12.5)[cb]{\scriptsize $\times$}
\Text(-4,-3)[cb]{\small $4$}
\Text(45,-18)[cb]{\small $1$}
\Text(45,12)[cb]{\small $2$}
\Text(12,19)[cb]{\small $8$}
\Text(28,19)[cb]{\small $16$}
\end{picture}
}
+
\scalebox{.7}{
\begin{picture}(55,30)(-7,-2)
\Line(0,0)(10,0)
\Line(40,-15)(28,-6)
\Line(40,15)(28,6)
\CArc(20,0)(10,143,37)
\Line(12,15)(12,6)
\Text(12,12.5)[cb]{\scriptsize $\times$}
\Line(28,15)(28,6)
\Text(28,12.5)[cb]{\scriptsize $\times$}
\Text(-4,-3)[cb]{\small $4$}
\Text(45,-18)[cb]{\small $2$}
\Text(45,12)[cb]{\small $1$}
\Text(12,19)[cb]{\small $8$}
\Text(28,19)[cb]{\small $16$}
\end{picture}
}
\;\;.
\nonumber
\eeq

\noindent
The tree diagrams have been drawn in such a way, that one can
easily identify the original sequence of the loop lines (called 
``loop flow"), starting from the external loop leg with tag number 
$2^L$ and ending with the external loop leg with tag number 
$2^{L+1}$. 
The selection rules must discard the redundant contributions and be 
simple enough to be translated from the diagrammatic representation 
into the language of off-shell currents.
Two rules accomplishing this are:

\begin{itemize}

\vspace{-.4cm}
\item [1)] 
The current containing the first external line enters the 
loop flow first.

\vspace{-.3cm}
\item [2)] 
The 3 currents containing the external legs with the 3 smallest 
binary numbers enter the loop flow in fixed order (for example 
given by the ascending order of the 3 binaries).
\end{itemize}

\vspace{-.35cm}
\noindent
Applying the first rule to our example, we discard 4 redundant 
diagrams:

\vspace{-.6cm}
\beq
\scalebox{.7}{
\begin{picture}(55,30)(-7,-2)
\Line(0,0)(10,0)
\Line(40,-15)(28,-6)
\Line(40,15)(28,6)
\CArc(20,0)(10,0,360)
\Text(-4,-3)[cb]{\small $1$}
\Text(45,-18)[cb]{\small $2$}
\Text(45,12)[cb]{\small $4$}
\end{picture}
}
\;\to\;\;
\scalebox{.7}{
\begin{picture}(55,30)(-7,-2)
\Line(0,0)(10,0)
\Line(40,-15)(28,-6)
\Line(40,15)(28,6)
\CArc(20,0)(10,143,37)
\Line(12,15)(12,6)
\Text(12,12.5)[cb]{\scriptsize $\times$}
\Line(28,15)(28,6)
\Text(28,12.5)[cb]{\scriptsize $\times$}
\Text(-4,-3)[cb]{\small $1$}
\Text(45,-18)[cb]{\small $2$}
\Text(45,12)[cb]{\small $4$}
\Text(12,19)[cb]{\small $8$}
\Text(28,19)[cb]{\small $16$}
\end{picture}
}
+
\scalebox{.7}{
\begin{picture}(55,30)(-7,-2)
\Line(0,0)(10,0)
\Line(40,-15)(28,-6)
\Line(40,15)(28,6)
\CArc(20,0)(10,143,37)
\Line(12,15)(12,6)
\Text(12,12.5)[cb]{\scriptsize $\times$}
\Line(28,15)(28,6)
\Text(28,12.5)[cb]{\scriptsize $\times$}
\Text(-4,-3)[cb]{\small $1$}
\Text(45,-18)[cb]{\small $4$}
\Text(45,12)[cb]{\small $2$}
\Text(12,19)[cb]{\small $8$}
\Text(28,19)[cb]{\small $16$}
\end{picture}
}
+
\scalebox{.7}{
\begin{picture}(55,30)(-7,-2)
\Line(0,0)(10,0)
\Line(40,-15)(28,-6)
\Line(40,15)(28,6)
\CArc(20,0)(10,143,37)
\Line(12,15)(12,6)
\Text(12,12.5)[cb]{\scriptsize $\times$}
\Line(28,15)(28,6)
\Text(28,12.5)[cb]{\scriptsize $\times$}
\Text(-4,-3)[cb]{\small $2$}
\Text(45,-18)[cb]{\small $1$}
\Text(45,12)[cb]{\small $4$}
\Text(12,19)[cb]{\small $8$}
\Text(28,19)[cb]{\small $16$}
\SetWidth{2}
\Line(0,20)(40,-20)
\Line(0,-20)(40,20)
\end{picture}
}
+
\scalebox{.7}{
\begin{picture}(55,30)(-7,-2)
\Line(0,0)(10,0)
\Line(40,-15)(28,-6)
\Line(40,15)(28,6)
\CArc(20,0)(10,143,37)
\Line(12,15)(12,6)
\Text(12,12.5)[cb]{\scriptsize $\times$}
\Line(28,15)(28,6)
\Text(28,12.5)[cb]{\scriptsize $\times$}
\Text(-4,-3)[cb]{\small $2$}
\Text(45,-18)[cb]{\small $4$}
\Text(45,12)[cb]{\small $1$}
\Text(12,19)[cb]{\small $8$}
\Text(28,19)[cb]{\small $16$}
\SetWidth{2}
\Line(0,20)(40,-20)
\Line(0,-20)(40,20)
\end{picture}
}
+
\scalebox{.7}{
\begin{picture}(55,30)(-7,-2)
\Line(0,0)(10,0)
\Line(40,-15)(28,-6)
\Line(40,15)(28,6)
\CArc(20,0)(10,143,37)
\Line(12,15)(12,6)
\Text(12,12.5)[cb]{\scriptsize $\times$}
\Line(28,15)(28,6)
\Text(28,12.5)[cb]{\scriptsize $\times$}
\Text(-4,-3)[cb]{\small $4$}
\Text(45,-18)[cb]{\small $1$}
\Text(45,12)[cb]{\small $2$}
\Text(12,19)[cb]{\small $8$}
\Text(28,19)[cb]{\small $16$}
\SetWidth{2}
\Line(0,20)(40,-20)
\Line(0,-20)(40,20)
\end{picture}
}
+
\scalebox{.7}{
\begin{picture}(55,30)(-7,-2)
\Line(0,0)(10,0)
\Line(40,-15)(28,-6)
\Line(40,15)(28,6)
\CArc(20,0)(10,143,37)
\Line(12,15)(12,6)
\Text(12,12.5)[cb]{\scriptsize $\times$}
\Line(28,15)(28,6)
\Text(28,12.5)[cb]{\scriptsize $\times$}
\Text(-4,-3)[cb]{\small $4$}
\Text(45,-18)[cb]{\small $2$}
\Text(45,12)[cb]{\small $1$}
\Text(12,19)[cb]{\small $8$}
\Text(28,19)[cb]{\small $16$}
\SetWidth{2}
\Line(0,20)(40,-20)
\Line(0,-20)(40,20)
\end{picture}
},
\nonumber
\eeq

\vspace{+.1cm}
\noindent
while the second rule excludes the last unwanted contribution

\vspace{-.3cm}
\beq
\scalebox{.7}{
\begin{picture}(55,30)(-7,-2)
\Line(0,0)(10,0)
\Line(40,-15)(28,-6)
\Line(40,15)(28,6)
\CArc(20,0)(10,0,360)
\Text(-4,-3)[cb]{\small $1$}
\Text(45,-18)[cb]{\small $2$}
\Text(45,12)[cb]{\small $4$}
\end{picture}
}
\;\to\;\;
\scalebox{.7}{
\begin{picture}(55,30)(-7,-2)
\Line(0,0)(10,0)
\Line(40,-15)(28,-6)
\Line(40,15)(28,6)
\CArc(20,0)(10,143,37)
\Line(12,15)(12,6)
\Text(12,12.5)[cb]{\scriptsize $\times$}
\Line(28,15)(28,6)
\Text(28,12.5)[cb]{\scriptsize $\times$}
\Text(-4,-3)[cb]{\small $1$}
\Text(45,-18)[cb]{\small $2$}
\Text(45,12)[cb]{\small $4$}
\Text(12,19)[cb]{\small $8$}
\Text(28,19)[cb]{\small $16$}
\end{picture}
}
+
\scalebox{.7}{
\begin{picture}(55,30)(-7,-2)
\Line(0,0)(10,0)
\Line(40,-15)(28,-6)
\Line(40,15)(28,6)
\CArc(20,0)(10,143,37)
\Line(12,15)(12,6)
\Text(12,12.5)[cb]{\scriptsize $\times$}
\Line(28,15)(28,6)
\Text(28,12.5)[cb]{\scriptsize $\times$}
\Text(-4,-3)[cb]{\small $1$}
\Text(45,-18)[cb]{\small $4$}
\Text(45,12)[cb]{\small $2$}
\Text(12,19)[cb]{\small $8$}
\Text(28,19)[cb]{\small $16$}
\SetWidth{2}
\Line(0,20)(40,-20)
\Line(0,-20)(40,20)
\end{picture}
}
\;\;.
\nonumber
\eeq

\vspace{+.2cm}
Having reduced the formal generation of the one-loop amplitude to the
generation of a set of tree-level processes, we can build the ``loop
off-shell currents" in a similar way as the tree-level currents, by 
means of a recursion relation:

\vspace{-.6cm}
\beq
\scalebox{.9}{
\begin{picture}(55,40)(-5,-2)
\Line(20,20)(20,10)
\Text(20.2,17.5)[cb]{\scriptsize $\times$}
\Text(13,20)[cb]{\scriptsize $P_0$} 
\Line(20,0)(40,0)
\GCirc(40,0){1.5}{0}
\Line(0,15)(20,0)
\Line(0,-15)(20,0)
\GCirc(20,0){10}{.8}
\DashCArc(20,0)(22,155,205){2}
\Text(-9,-3)[cb]{$n$} 
\Text(42,5)[cb]{$P$} 
\end{picture}
}
\quad
=
\quad
\sum_{\{i\},\{j\}}^{i+j=n}\;
\sum_{P_i,P_j}\;
\scalebox{.9}{
\begin{picture}(65,40)(-10,-2)
\Line(15,30)(15,22.5)
\Text(15.2,27.5)[cb]{\scriptsize $\times$}
\Text(8,30)[cb]{\scriptsize $P_0$} 
\Line(0,25)(15,15)
\Line(0,5)(15,15)
\Line(15,15)(35,0)
\GCirc(15,15){7.5}{.8}
\DashCArc(15,15)(16,160,200){2}
\Text(31,8)[cb]{\scriptsize $P_{\!i}$}
\Text(-6,12)[cb]{$i$} 
\Line(0,-5)(15,-15)
\Line(0,-25)(15,-15)
\Line(15,-15)(35,0)
\GCirc(15,-15){7.5}{.8}
\DashCArc(15,-15)(16,160,200){2}
\Text(31,-17)[cb]{\scriptsize $P_{\!j}$}
\Text(-6,-18)[cb]{$j$} 
\SetWidth{1.5}
\Line(35,0)(50,0)
\SetWidth{.5}
\BBoxc(34,0)(5,5)
\GCirc(50,0){1.5}{0}
\Text(52,5)[cb]{$P$} 
\end{picture}
}
\quad
+
\;
\sum_{\{i\},\{j\},\{k\}}^{i+j+k=n}\;
\sum_{P_{\!i},P_{\!j},P_{\!k}}\;
\scalebox{.9}{
\begin{picture}(70,40)(-10,-2)
\Line(15,40)(15,32.5)
\Text(15.2,37.5)[cb]{\scriptsize $\times$}
\Text(8,40)[cb]{\scriptsize $P_0$} 
\Line(0,35)(15,25)
\Line(0,15)(15,25)
\Line(15,25)(45,0)
\GCirc(15,25){7.5}{.8}
\DashCArc(15,25)(16,160,200){2}
\Text(-6,22)[cb]{$i$} 
\Text(31,18)[cb]{\scriptsize $P_{\!i}$}
\Line(0,10)(15,0)
\Line(0,-10)(15,0)
\Line(15,0)(45,0)
\GCirc(15,0){7.5}{.8}
\DashCArc(15,0)(16,160,200){2}
\Text(-6,-3)[cb]{$j$} 
\Text(27,1)[cb]{\scriptsize $P_{\!j}$}
\Line(0,-15)(15,-25)
\Line(0,-35)(15,-25)
\Line(15,-25)(45,0)
\GCirc(15,-25){7.5}{.8}
\DashCArc(15,-25)(16,160,200){2}
\Text(-6,-28)[cb]{$k$} 
\Text(31,-27)[cb]{\scriptsize $P_{\!k}$}
\SetWidth{1.5}
\Line(45,0)(60,0)
\SetWidth{.5}
\BBoxc(43,0)(5,5)
\GCirc(60,0){1.5}{0}.
\Text(60,5)[cb]{$P$}
\end{picture}
}
\;\;\;.
\label{recursive loop}
\eeq

\vspace{.4cm}
\noindent
Formally the relation is the same as at tree level, with the 
additional leg $P_0$ (at loop level the particle $\bar{P}_0$ is 
chosen as last). 
The main difference comes actually form the fact that every loop 
current contains a dependence on the integration momentum $q$, 
generated by the Feynman rules for the vertex and the propagator.
Working in the 't Hooft--Feynman gauge in the SM, the $q$-dependence
of (vertex)$\times$(propagator) takes the form

\vspace{-.4cm}
\beq
(\mbox{vertex}) \;\times\; (\mbox{propagator}) 
\,=\,
\frac{a_\mu q^\mu + b}{(q+p)^2-m^2}
\;.
\label{loop vertexprop}
\eeq

\noindent
This $q$-dependence determines through repeated application of the 
recursion relation the structure of the loop currents, 
which after $k$ steps gets a $q$-dependent structure of the type:

\vspace{-.3cm}
\beq
\mbox{loop current}\;(q)
\;=\;
\sum_{r\,=\,0}^k\;
a_{\mu_1\cdots\mu_r}^{(k,r)}\;
\frac{q^{\mu_1}\cdots q^{\mu_r}}{\prod_{h=1}^k[(q+p_h)^2-m_h^2]}
\label{loop current (q)}
\;,
\eeq

\noindent
where $k$ is the number of $q$-dependent propagators and $r$ is the 
rank of each term of the linear combination.
The $q$-dependence of the loop currents in \refeq{loop current (q)} 
has the same structure as the one-loop amplitude of 
\refeq{tensor splitting}. 
In particular the coefficients $a_{\mu_1\cdots\mu_r}^{(k,r)}$ that we 
get at the last step of the procedure coincide with the TCs 
$c_{\mu_1\cdots\mu_{r_t}}^{(t)}$. 
Therefore we use the recursive relation in \refeq{recursive loop} to 
compute at each step the coefficients $a_{\mu_1\cdots\mu_r}^{(k,r)}$ of 
the loop current, keeping track of the corresponding loop 
propagators.
This is done by defining the $q$-independent loop off-shell 
current

\vspace{-1.45cm}
\beq
w_{i_k}(P,{\cal C},B,\{p_1,\dots,p_k\},\{m_1,\dots,m_k\})
\;=\;\,
\begin{picture}(55,70)(-20,-14)
\Line(0,18)(20,10)
\Line(0, 2)(20,10)
\Line(0, -4)(20,-12)
\Line(0,-20)(20,-12)
\Line(0,-26)(20,-34)
\Line(0,-42)(20,-34)
\DashCArc(20,10)(22,165,195){2}
\DashCArc(20,-12)(22,165,195){2}
\DashCArc(20,-34)(22,165,195){2}
\DashCArc(200,-20)(210,170,186){2}
\Text(-16,-6)[cb]{\scriptsize $n$} 
\Line(20,22)(20,-15)
\Text(40,-5)[cb]{\scriptsize $(p_1,m_1)$} 
\DashLine(20,-12)(20,-34){1.5}
\Text(20,19.5)[cb]{\scriptsize $\times$}
\Line(20,-34)(35,-34)
\GCirc(20, 10){5}{.8}
\GCirc(20,-12){5}{.8}
\GCirc(20,-34){5}{.8}
\GCirc(35,-34){1.5}{0}
\Text(43,-37)[cb]{\scriptsize $P$} 
\end{picture}
\qquad\;,
\label{loop current}
\eeq

\vspace{.75cm}
\noindent
where the sequences of off-set momenta $\{p_1,\dots,p_k\}$ and of 
masses $\{m_1,\dots,m_k\}$ of the propagators link each $w_{i_k}$
with its $q$ structure and at the end of the procedure allow to 
relate each TC to its TI. 
The tensorial index $i_k$ of the loop current $w_{i_k}$ is a 
short-hand notation to express the coefficients 
$a_{\mu_1\cdots\mu_r}^{(k,r)}$ for all values of $r$ and 
$\mu_1\cdots\mu_r$:

\vspace{-.65cm}
\beq
i_k\! = 0 \;\to\; w_{i_k}\! = a^{(k,0)}\!\!, \quad\;\;
i_k\! = 1,\dots,4  \;\to\; w_{i_k}\! = a^{(k,1)}_{\mu_1}\!\!, \quad\;\;
i_k\! = 5,\dots,14 \;\to\; w_{i_k}\! = a^{(k,2)}_{\{\mu_1\mu_2\}}, \quad
\dots\;,
\eeq

\vspace{-.15cm}
\noindent
where we considered that only symmetric combinations of the indices 
$\mu_1,\dots,\mu_r$ give non-vanishing contributions.

In order to apply the recursive relation in \refeq{recursive loop}, 
it remains to understand which single-particle currents have to be 
associated with the new particles $P_0$ and $\bar{P}_0$.
This is done by introducing for fermions and vector bosons a suitable 
set of spinors $\psi_i$ and polarization vectors {$\epsilon_i^\mu$}, 
in such a way that the contraction originally contained in the loop 
diagrams can be reproduced:

\vspace{-.7cm}
\beq
(\psi_i)_{\alpha}\! = (\bar{\psi}_i)_{\alpha}\! = \delta_{i\alpha}
\quad\mbox{with}\quad
\sum_{i=1}^4(\bar{\psi}_i)_{\alpha}(\psi_i)_{\beta}\! = \delta_{\alpha\beta},
\qquad\quad
\epsilon_i^\mu\!\! = \delta_i^\mu
\quad\mbox{with}\quad
\sum_{i=1}^4\!\epsilon_i^\mu\!\epsilon_i^\nu\! = \delta^{\mu\nu},
\label{cutspinors}
\eeq

\vspace{-.2cm}
\noindent
where $i$ denotes the ``polarization'', and $\alpha,\beta$ and
$\mu,\nu$ are spinor and Lorentz indices, respectively.

\section{Treatment of colour}
\label{colour}

\vspace{-.2cm}
We use the colour-flow representation introduced in 
\citere{Kanaki:2000ms}, where the conventional 8 gluon fields 
$A_{\mu}^{a}$ are replaced by a $3\times 3$ matrix 
$({\cal A}_{\mu})^{i}_{j} = 
\frac{1}{\sqrt{2}}\,A_{\mu}^{a}(\lambda^{a})^{i}_{j}$ with the trace
condition $\sum_{i}\,({\cal A}_{\mu})^{i}_{i} = 0$.  Quarks and
antiquarks maintain the usual colour index $i=1,{2},3$, while gluons
get a pair of indices $i,j=1,{2},3$.
The Feynman rules in this representation are in practice obtained by 
multiplying all gluon lines by $(\lambda^{a})^{i}_{j}$ in the 
standard Feynman rules.
Using the algebra of the Gell-Mann matrices, the colour part of the 
new Feynman rules consists simply in combinations of Kronecker 
$\delta$s. 
For example one gets:

\vspace{-.8cm}
\beqar
\!\!\!\!
&&
\!\!
\begin{picture}(40,10)(-5,-2)
\ArrowLine(0,0)(25,0)
\Text(-4,-4)[cb]{{\scriptsize $j$}}
\Text(30,-2)[cb]{{\scriptsize $i$}}
\Text(13,-12)[cb]{\scriptsize $p$}
\end{picture}
=
\delta^{i}_{j}
\times
\frac{\ri(\ps+m)}{p^2-m^2},
\qquad\;\;
\begin{picture}(57,10)(-20,-2)
\Gluon(0,0)(25,0){2.5}{4}
\Text(-5,1)[cb]{\scriptsize $i_1$}
\Text(-5,-9)[cb]{\scriptsize $j_1$}
\Text(31,-9)[cb]{\scriptsize $i_2$}
\Text(31,1)[cb]{\scriptsize $j_2$}
\Text(4,3)[cb]{\scriptsize $\mu$}
\Text(22,4)[cb]{\scriptsize $\nu$}
\Text(13,-12)[cb]{\scriptsize $p$}
\end{picture}
=
\begin{picture}(50,10)(-12,-2)
\ArrowLine(25,3)(0,3)
\ArrowLine(0,-3)(25,-3)
\Text(-6,1)[cb]{\scriptsize $i_1$}
\Text(-6,-9)[cb]{\scriptsize $j_1$}
\Text(32,-9)[cb]{\scriptsize $i_2$}
\Text(32,1)[cb]{\scriptsize $j_2$}
\end{picture}
\times
\frac{-\,\ri\,g_{\mu\nu}}{p^2}
\;
=
\;
\delta^{i_1}_{j_2}\delta^{i_2}_{j_1}\,
\frac{-\,\ri\,g_{\mu\nu}}{p^2},
\nonumber\\[-.4cm]
&&
\!\!\!\!
\scalebox{0.8}{
\begin{picture}(63,50)(-8,-2)
\Gluon(20,0)(43,0){2.5}{4}
\ArrowLine(20,0)(0,20)            
\ArrowLine(0,-20)(20,0)
\Vertex(20,0){2}
\Text(-5,19)[cb]{$i_1$} 
\Text(-5,-26)[cb]{$j_2$}
\Text(51,-9)[cb]{$i_3$}
\Text(51,1)[cb]{$j_3$}
\Text(39,3)[cb]{$\mu$}
\end{picture}
}
\;=
\,
\left(
\scalebox{0.8}{
\begin{picture}(59,35)(-8,-2)
\ArrowLine(17,3)(0,20)     
\ArrowLine(40,3)(17,3)
\ArrowLine(0,-20)(17,-3)
\ArrowLine(17,-3)(40,-3)
\Text(-5,19)[cb]{$i_1$} 
\Text(-5,-24)[cb]{$j_2$}
\Text(48,-9)[cb]{$i_3$}
\Text(48,1)[cb]{$j_3$}
\end{picture}
}
 - \,\frac{1}{\NC}
\scalebox{0.8}{
\begin{picture}(59,35)(-8,-2)
\ArrowLine(17,3)(0,20)
\ArrowLine(0,-20)(17,-3)
\CArc(14,0)(4.2,-45,45)
\DashLine(18.2,0)(30.8,0){3}
\CArc(35,0)(3,90,270)
\ArrowLine(43,3)(35,3)
\ArrowLine(35,-3)(43,-3)
\Text(-5,19)[cb]{$i_1$} 
\Text(-5,-24)[cb]{$j_2$}
\Text(51,-9)[cb]{$i_3$}
\Text(51,1)[cb]{$j_3$}
\end{picture}
}
\right)
\!\!\times\!
\frac{\ri\,\gs}{\sqrt{2}}\,\gamma^\mu
\;=
\,
\left(\!
\delta^{i_1}_{j_3}\delta^{i_3}_{j_2}
- \frac{1}{\NC}\delta^{i_1}_{j_2}\delta^{i_3}_{j_3}
\!\right)\!
\frac{\ri\,\gs}{\sqrt{2}}\,\gamma^\mu\!.
\qquad\;\;
\label{colourflow rules}
\eeqar

\noindent
As a consequence, in the colour-flow representation the colour 
structure of the amplitude for a process with $k$ external gluons and 
$m$ external quark-antiquark pairs can be written as a
linear combination of all possible products of Kronecker $\delta$s:

\vspace{-.3cm}
\beq
{\cal A}^{\alpha_1,\dots,\alpha_n}_{\beta_1,\dots,\beta_n} = 
\sum_{P(\beta_1,\dots,\beta_n)} 
\delta^{\alpha_1}_{\beta_1}\!\cdots\delta^{\alpha_n}_{\beta_n}\,
{\cal A}_P,
\qquad
n = k+m,
\label{colourflow amplitude}
\eeq

\noindent
where we do not pose any restriction on the permutations 
$P(\beta_1,\dots,\beta_n)$ to be considered.
At this point basically two different strategies are possible for the 
computation of the sum over colours of the squared amplitude.
In a framework based on colour-dressed amplitudes, all $N_c^{2n}$ 
amplitudes ${\cal A}^{\alpha_1,\dots,\alpha_n}_{\beta_1,\dots,\beta_n}$ 
are computed (i.e. for all possible colour indices 
$\alpha_1,\dots,\alpha_n,\beta_1,\dots,\beta_n$) and the summed squared 
amplitude is given by

\vspace{-.3cm}
\beq
{\cal A}^2
\;=
\sum_{\alpha_1\dots\alpha_n,\beta_1,\dots,\beta_n}
({\cal A}\,^{\alpha_1\cdots\alpha_n}_{\beta_1\cdots\beta_n})^*
{\cal A}\,^{\alpha_1\cdots\alpha_n}_{\beta_1\cdots\beta_n} .
\eeq

\noindent
Alternatively one can decide to use the decomposition of 
\refeq{colourflow amplitude} and work directly with the $n!$ 
``structure-dressed'' amplitudes ${\cal A}_P$ of the 
{\it colour structures} 
$\delta^{\alpha_1}_{\beta_1}\!\cdots\delta^{\alpha_n}_{\beta_n}$. 
In this approach the summed squared amplitude is obtained by summing 
over the permutations according to

\vspace{-.4cm}
\beq
{\cal A}^2
\;=\;
\sum_{P,P'}\,
{\cal A}_P^*\;C_{PP'}\;{\cal A}_{P'},
\eeq

\vspace{-.2cm}
\noindent
where the coefficients $C_{PP'}$ are trivially calculated (and simply 
given by positive powers of $N_c$).
This second strategy reduces the number of amplitudes to compute and 
is essentially the starting point of the known approach based on 
colour-ordered amplitudes (which aims to reduce the number of 
amplitudes even more). 
Our implementation is slightly different. 
We assign a colour structure (labeled by the index ${\cal C}$) to 
each off-shell current and derive rules to build at each step of the 
recursion procedure the colour structure of the outgoing current from 
the colour structures of the incoming ones. 
In this way the colour structure of the amplitude is automatically 
generated, all structure-dressed amplitudes ${\cal A}_P$ are computed 
simultaneously in {\it one} recursive procedure, and 
structure-dressed currents which contribute at intermediate steps to 
several ${\cal A}_P$ are computed just once.
Moreover one can further optimize the computation of currents 
which only differ by the colour structure, by recomputing just their 
colour coefficient instead of the whole current. 

\section {Features of \recola\ }
\label{features}

\vspace{-.2cm}
The code \recola\ is structured in two parts: the generation of the 
recursion procedure (to be run once) and the computation of the 
currents (to be run at each phase-space point). 
In the generation procedure, the skeleton of the recursion procedure 
is created, assigning an index to each current and each branch. 
The list of all needed TIs is also generated at this point.
Currents differing just by the colour structures are identified and 
their colour coefficients are computed.
In the second part all the TIs are computed (calling external 
libraries) as well as all currents for all polarizations, 
according to the skeleton derived in the generation part. 
Finally the TCs (i.e. the final currents) are contracted with the TIs 
to give the structure-dressed amplitudes ${\cal A}_P$, from which the 
squared amplitude summed over colour and polarization is calculated.

\recola\ allows the computation of tree and one-loop amplitudes in 
the full SM, including all counterterms~\cite{Denner:1991kt} and 
rational parts of type R2\footnote{Since the loop currents are 
computed numerically, their indices are strictly 4-dimensional and 
the R2 part must be added separately.}~\cite{Ossola:2008xq}.
A consistent treatment of unstable particles is provided by using 
the complex-mass scheme~\cite{Denner:2005fg}. 
On-shell renormalization is used for the EW sector, while the 
strong coupling constant is renormalized in the $\overline{\rm MS}$
scheme. 
The ultraviolet pole structure of one-loop amplitudes can be accessed 
numerically and for soft/collinear divergencies either dimensional or 
mass regularisation can be employed. 
For a given process at tree or one-loop level \recola\ allows to 
single out arbitrary powers in the strong coupling constant. 
In addition \recola\ calculates all colour- and spin-correlated 
amplitudes needed for the application of the dipole subtraction 
method~\cite{Catani:1996vz}.
The calculation of one-loop amplitudes relies on the external 
computation of the tensor integrals which is done with the COLLIER 
library~\cite{Denner:2002ii}.


In addition to an efficient treatment of colour, 
also the sum over helicity configurations has been optimized by 
avoiding recalculation of identical currents appearing in different 
helicity configurations together with the use of helicity 
conservation for massless fermions in the SM. 
The resulting code requires negligible amount of memory for executables, 
object files and libraries, while the RAM needed does not exceed 2 
Gbyte even for complicated processes. 
For the sample processes in Table 1, the CPU time needed in the 
generation is of the order of some seconds, while the computation 
of the TCs takes at most few hundreds of milliseconds, usually of the 
same order of magnitude as the computation of the TIs with COLLIER. 

\begin{table}
\begin{center}
\begin{tabular}{|c|c|cc|cc|cc|}
\hline
Process & $t_{\rm TIs}$ & 
$t_{\rm gen}$ & $t_{\rm TCs}$ & 
$t_{\rm gen}$ & $t_{\rm TCs}$ & 
$t_{\rm gen}$ & $t_{\rm TCs}$ 
\\[-.1cm]
& \hspace{-.1cm}(COLLIER)\hspace{-.1cm} &
\multicolumn{2}{|c|}{(single helicity)} & 
\multicolumn{2}{|c|}{(partial hel. sum)} & 
\multicolumn{2}{|c|}{(helicity sum)} \\
\hline\hline
$u \bar{u} \to W^+ W^- g$ & 2.8 ms &
0.3 s & 0.6 ms & 
0.4 s & 1.3 ms & 
1.6 s & 9.8 ms
\\[-.15cm]
(QCD) & &
\multicolumn{2}{|c|}{(hel: - + - + -)} & 
\multicolumn{2}{|c|}{(hel: S S - + S)} & 
\multicolumn{2}{|c|}{(hel: S S S S S)} \\
\hline
$u \bar{d} \to W^+ g\, g\, g$ & 130 ms &
14 s &  14 ms & 
25 s &  76 ms & 
52 s & 221 ms
\\[-.15cm]
(QCD) & &
\multicolumn{2}{|c|}{(hel: - + - - - -)} & 
\multicolumn{2}{|c|}{(hel: S S - S S S)} & 
\multicolumn{2}{|c|}{(hel: S S S S S S)\hspace{-.1cm}} \\
\hline
$u g \to u\, g\, Z$ & 8.2 ms &
0.5 s &  1.4 ms & 
1.0 s &  6.7 ms & 
2.2 s & 20.2 ms
\\[-.15cm]
(EW) & &
\multicolumn{2}{|c|}{(hel: - - - - -)} & 
\multicolumn{2}{|c|}{(hel: S S S S -)} & 
\multicolumn{2}{|c|}{(hel: S S S S S)} \\
\hline
$u g \to u\, g\, \tau^- \tau^+$ & 28 ms &
1.3 s &  2.5 ms & 
2.0 s & 14.2 ms & 
3.8 s & 29.0 ms
\\[-.15cm]
(EW) & &
\multicolumn{2}{|c|}{(hel: - - - - - +)} & 
\multicolumn{2}{|c|}{(hel: S S S S - +)} & 
\multicolumn{2}{|c|}{(hel: S S S S S S)\hspace{-.2cm}} \\
\hline
\end{tabular}
\end{center}
\vspace{-.5cm}
\caption{CPU time needed by \recola\ for the computation of sample 
processes with different helicity configurations (the ``S'' means 
that the sum over helicities has been taken for the corresponding 
particle).
The computation has been performed with a processor Intel(R) 
Core(TM) i5-2400 CPU @ 3.10GHz.}
\label{cpu}
\end{table}

\section {Conclusion}

\vspace{-.2cm}
We have presented \recola, an efficient code based on a one-loop 
generalization of Dyson-Schwinger recursion relations, which 
automatizes the computation of EW and QCD processes with elementary 
particles in the SM at NLO. 
The code has been used for the computation of EW corrections to 
the process $p p \to Z + 2j$ \cite{Actis:2012qn,Denner:radcor13}.

\section*{Acknowledgments}

\vspace{-.2cm}
This work was supported in part by the Deutsche 
Forschungsgemeinschaft (DFG) under reference number DE~623/2-1.

\end{document}